# On the Configuration of EM Waves and Bosons

Surendra Mund*
*Student,* Department of Physics, Central University of Rajasthan, Bandarsindri, India

*Abstract*: This is the first expression of my thoughts and my experiments with Nature about the mathematical description of the Universe. The theories about our surrounding Nature became popular from our ancient civilizations and may be from the starting of life. So, by coming straight forward to the point light had been concerned as the most remarkable thing from the starting time of life and here my first paper is based on the particular physical property of Universe. Here, I am giving the particular principle of light leaving behind the last 5,000 years to the civilizations and the mathematical and philosophical aspects of the particular type of thinkers. From the thinkers of India and the westerns the light had been described as the main component of human and the parallel living lives on our planet, but the main target of the paper is not to show or prove the history about the mankind. In this paper we will look at the theories about light then my challenges and the solutions about the particular principle will be obtained.

## 1. Introduction

First Romer (the astronomer) proved that light have a particular speed or in another sense it does not have infinite speed. After that the corpuscular theory by Newton and wave theory by Christian Huygens were given but for 300 years only the theory by Christian Huygens only accepted because there were some remarkable experiments in favor of it like light shows Reflection, Interference, Dispersion, Refraction, Scattering, Polarization etc. but in 1901 Max Planck used the ideology of Newton to explain the Blackbody Radiation and Correspondingly Albert Einstein explained the Photoelectric effect by the same ideology. So, first I will give a brief introduction to my explanation about light and then I will prove all the concerned experiments by logic , then we will go further to the wave- particle duality and the explanation of the Cherenkov Radiation and some another beautiful mathematical aspects of my thoughts about light.

## 2. My Generalization of Some Former Principles about Light

I am starting from the geometry and coupling of EM waves and the principles about it that how light do couple with the surrounding scalar field and what are the effects of the density of scalar field on the propagation of light and when it converted into particle and when it behaves like wave ? These all questions have a particular sense in our above discussion. If you think about the creation of light particles then you will reach at a geometrical view of light from electrons (Or interruption into the flow of electron's surrounding scalar field). Let me explain the above thinking trend by a Diagram-

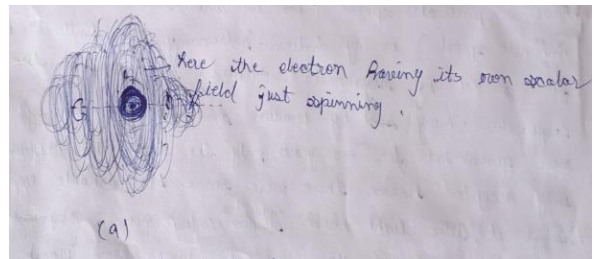

Fig.1.Geometrical Representation-1

If there is an outside effect on electron's surrounding scalar field occur then some part of it will be removed from this particle and we will get the situation-

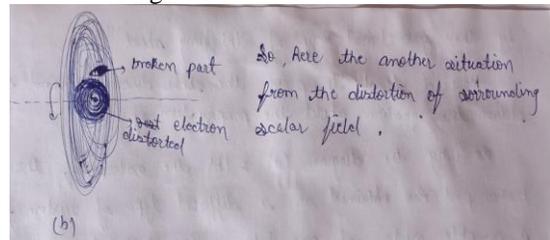

Fig.2. Geometrical Representation-2

So, from here the broken part forms another type of geometry and gets the spin of the particle and the surrounding scalar field and then propagates into space (or Universal Scalar Field at that particular epoch from where the particle created).
So, here the broken part has full coupling or partially full coupling with the surrounding scalar field and we obtain the geometry of the broken part like this-

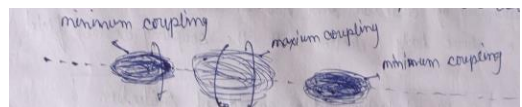

Fig.3.Configuration of Broken Part

When it approaches to the maximum coupling the broken part starts repulsing and then the particle reverse the configuration and comes to its initial state of formation and then again goes to maximum coupling. Somewhere and in some situations the maximum coupling state show its effect and in other situations the minimum coupling state shows its effect and the corresponding theories arise by looking on the effects of one type (wave and particle nature of light). So, some of the former observations like light does show reflection, refraction, interference, dispersion, polarization, scattering, photo electric effect, Doppler effect and Compton scattering can be explained by the geometrical way of thinking about it.

---

*Corresponding author: 2018imsph009@curaj.ac.in

Now I tend to prove the above discussion and may obtain the equations about the propagation of light.

So, by the diagram (a) & (b) the nature of the broken part has obtained as a different type of system from a complete system of one type. This included the distortion of scalar field or the dynamics of particle has been affected by that distortion of scalar or the dynamics of particle has been affected by that distortion and changed the motion of system. By change into the motion of a complete particle (particle with its surrounding scalar field) I tend to connect this change in motion into the various parameters of motion and propagation of the broken part. Clear from the above discussion the value of the broken part (or the quantity of the broken part) depends upon the quantity of distortion and how the distortion implemented on the above body (electron with the coupling scalar field).

$$\text{Quantity of broken part} \propto \text{quantity of Distortion}$$

Let's represent the quantity of broken part ($\Psi_b$) and quantity of distortion ($\delta_D$)-

$$\Psi_b \propto \delta_D \quad (1)$$

The quantity of distortion equals the change in motion of the system or the electron with the outer affected scalar field.

$$\delta_D \propto \text{Change in motion of system}$$

Here the motion depends upon the scalar field density on that epoch of universe and basically on the creation state of the body on the universal epoch.

$$\text{Quantity of motion} \propto \text{Scalar field density of universe and Nature of scalar field}$$

Here represent the quantity of motion by $Q_M$ and scalar field density by $\varphi_u$ (at the particular epoch of universe).

$$Q_M \propto \Phi u \quad (3)$$

Or the speed of propagation of broken part depend upon the coupling constant –

$$v_{p_b} \propto \epsilon \quad (4)$$

{here $\epsilon$ = coupling constant & $v_p$ = speed of propagation}

Here speed can be generalized as how much scalar field does covered by the broken part from maximum coupling state to minimum coupling state in universal scalar field at the particular epoch.

$$v_{p_b} = (\Phi u)_{covered}/\Delta \tau_{\varphi max-min} \quad (5)$$

{Here $\Delta \tau_{\varphi max-min}$ is the time quantity}

Here the coupling constant of the broken particles can vary with respect to change in the density of scalar field or space time (in terms of Albert).

$$(v_p)_b \propto \Phi u$$

$$(6)$$

So, from the above generalization we can conclude some important facts like the speed of light (broken particle) should be different in different phases of universe or speed of light (c) should not be a universal constant. From the another point of view space-time have a new definition that it is a flow of universal scalar field or for a particular system its space-time depends upon the evolution of its scalar field and flow of particular one.

So, from the above conclusions we might have some basic queries like what should be the different types of scalar fields in existing universe and how they behave in different physical conditions?

But first I tend to prove here the above Geometrical Aspects by proving the different properties of broken particle (or light)-

"Light shows reflection and refraction when it collides with a different phase or medium in sense like some solids and fluids." So, from the above observation about light we configure a picture about what is happening-

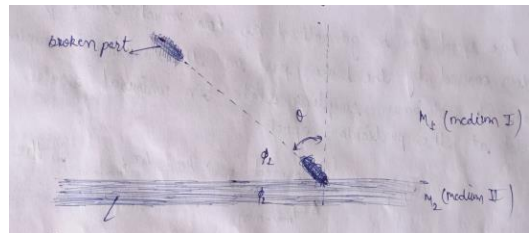

Fig.4.Refraction of Broken Part

Here the second medium have change into the density of scalar field and having some another type of coupling scalar field and here might exist some another type of central systems. When the broken part merge into the another type of coupling medium from $\varphi_1$ (from where the broken part was propagating) by an angle $\theta$ then this type of situation arises-

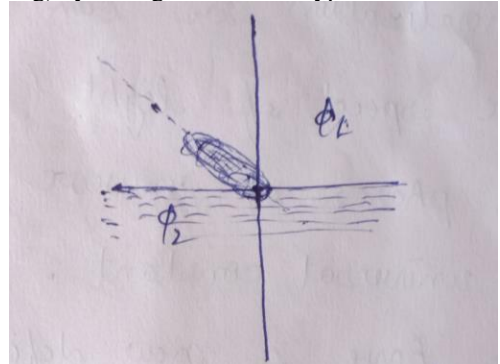

Fig.5.Closure look at Refraction Process

Here two conditions are applied to the broken part – the first one when $\varphi_1$ have more coupling then $\varphi_2$ and the second one is reversed from the first one. If $\varphi_1$ have the coupling constant $\epsilon_1$ and $\varphi_2$ have the coupling constant $\epsilon_2$ with the broken part then-

Condition 1:- if ($\epsilon_1 > \epsilon_2$) then the geometrical propagation of the broken part will be like the below diagram-

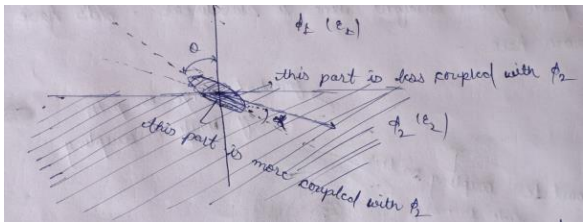
Fig.6.Refraction by entering in less coupling Scalar Field

So, the broken part will change its direction of propagation when it will enter into another type of coupling field. From the perspective of φ$_2$ the broken part will change its direction by an angle of α because of the coupling constant of φ$_2$ is different from the coupling constant of φ$_1$.

Condition 2:- if ($\epsilon_1 < \epsilon_2$) then-

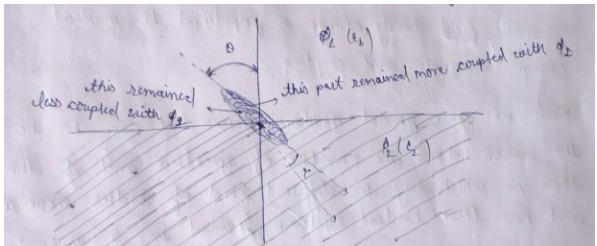
Fig.7.Refraction by entering in more coupling Scalar Field

Here in this condition the broken part change its path of propagation by α again. So, the above effect directly depends upon the angle of entrance θ. If θ will increase then the effect of coupling with another scalar field will be measured more and α (the angle of distortion) will be increased.

Now, you may think about the reason behind the bending of light by some different types of central systems like solar systems. If the density of scalar field (gradually increasing or decreasing) then the broken particle (or light) should change its path from straight line to curvilinear. Let's have a diagram here-

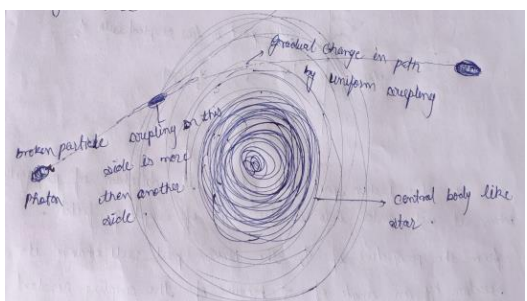
Fig.8.Bending of Broken Part by Gradual Variation in Scalar Field

$$\Delta \phi_d \propto \frac{\Delta \theta}{\Delta \tau} \quad (7)$$

Scalar field density change is proportional to gradual change in angle and coupling. From here some general questions arises like can this be the reason behind the curvature of spacetime and if this is then in which way the different kind of bodies move around and how these central systems formed in universe? By not concerning about the solutions of these questions here I only tend to prove other properties of the broken particle like diffraction by one slit and interference of photons by two slits. Here let's suppose the above geometrical trend in diffraction by one slit of the broken particle-

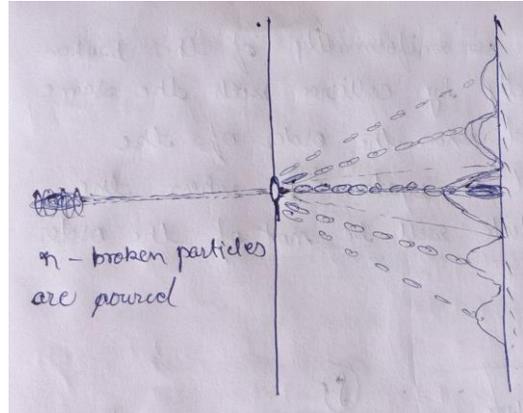
Fig.9.Differaction of Broken Part by One Slit

Here on interacting with a different medium on different ranges the broken particles are bent and observed on diffraction slit but on the few points of the film (which is solid and formed by different types of central systems) and here some friction will change the composition of the broken particle.

$$\Delta \Psi_b \propto f \quad (8)$$

Here $\Psi_b$ is the quantity of broken part.

Here change in the quantity of broken part will be converted into scalar field and in other words there will be some particle-scalar field transformation (or $\phi$-$\psi$ transformation) will occur.

Now come to the another explanation of interference by two slits in the same geometrical trend-

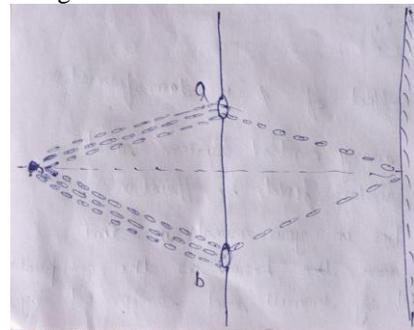
Fig.10.Two Slit Experiment Representation

Now if the maximum expansion or coupling state equals to the first medium (spacetime) or scalar field density on the position 'a' then by the above coupling phenomena we can conclude the bend in path of broken particle.

So, here discreteness and uniform nature of the broken particle will be shown respectively by collision with the same level quantity of motion and when the order of the quantity of motion is not same in system, and then the density of broken particle will be not of the order of the interaction place.

If
$$\phi_b \cong \phi_p \quad (9)$$

$\phi_p$ = scalar field density where the broken particle has changed the path or some $\phi$-$\psi$ transformation happened. Here in the situation (9) will behave discretely.

Or

$$\phi_b \neq \phi_p \tag{10}$$

In this case the broken particle will behave uniformly and there will be some change in scalar field or the particle will couple with the scalar field more from one side and change its path from straight line.

Now I tend to represent the polarization geometrically-

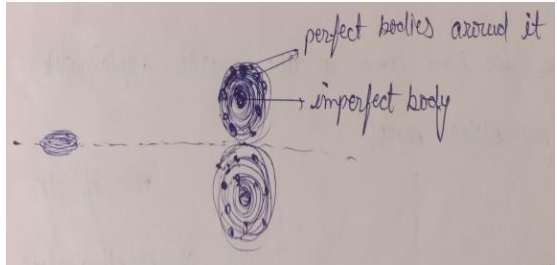

Fig.11.Polarization of Broken Part

Here we can easily see the effect of $\phi$-ψ transformation or the broken part couple with many central systems and gradually converting itself into scalar field by friction.
From here one question should be in your mind that if the broken part known as the combination of electric and magnetic field by J.C. Maxwell and confirmed by the experiment of Hertz, then what should be its new definition in terms of scalar field?

Here we are to just think about the geometry of the so-called broken particle-

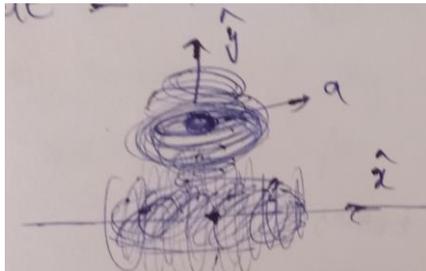

Fig.12.Geometrical Representation of Magnetic Effect

Here the entire scalar field is rotating in the direction of rotation of the broken particle. So, here two effects of the rotation arise. Suppose an another spinning particle come in reach of broken particle like particle (a) then this will feel coupling in another type and interrupt with the broken particle in the flow. But in another situation when the particle have the direction of spin in the direction of propagation, then-

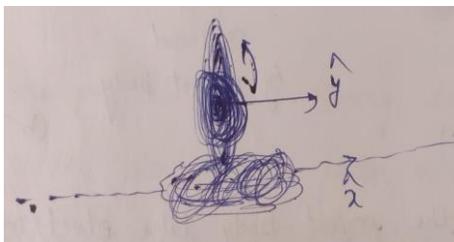

Fig.13. Geometrical Representation of Electric Effect

But in another situation the particle will feel another type of coupling with the broken particle. The first one is known as the magnetic effect and the second one is electric effect.

From the above discussion have some basic queries like electric and magnetic fields are connected by the geometrical configuration of particles and their interaction with the scalar field and how the change is generated or what is the basic definition of charge? Some points we can conclude from the above generalization like charge depends upon the scalar field of a body and its interaction with the scalar field when the field is rotating. Scalar field coupling with another body then in which way the motion of body and spin is affected will be discussed when the basic proofs and geometrical representations will be completed. Now I tend to give the geometrical generalization of photoelectric effect and Compton scattering. First drawing the photoelectric effect in the geometrical way-

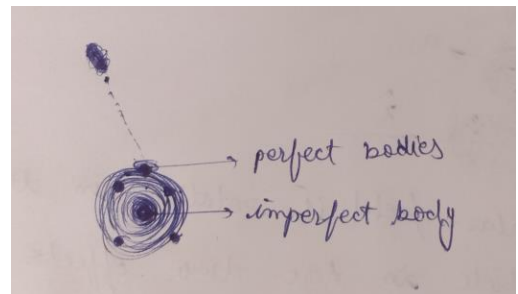

Fig.14. Geometrical Representation of Photo-Electric Effect

Here the broken part again coupled with electron, so the scalar field will be increased in this state when the electron is spinning on its particular axis. Let's see what is happening-

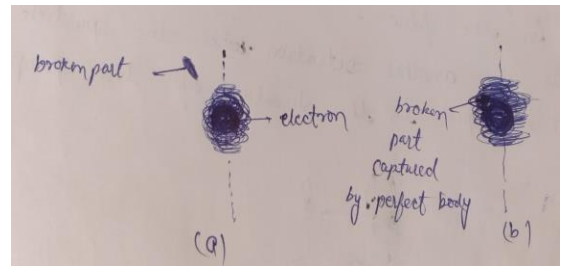

Fig.15.Broken Part Capturing Process

Here in situation (a) the perfect body like electron is spinning around a particular axis and the broken part couples in situation (b) then there is an instant increase seems in the coupling scalar field and increase in the quantity of the body and its coupling scalar field. If this coupling scalar field increase is such that electron or perfect body can be out from the central system then the broken part will be captured perfectly for a particular size of broken part. So, this effect is valid for a particular size of broken part not for the below size of broken parts because for the low sizes of broken parts there will be increase in the quantity of body and coupling scalar field but that will be not sufficient for taking out to the body from the central system. So, here should be a state after the capture of broken part which decides that the perfect body (like electron) will not be in central system or remaining on certain position after the capture will be the reason for the instability, so then the perfect body will release the broken

> *At minimum coupling state the broken part reverse the spin and the broken part starts motion in another direction but at the maximum coupling state the broken part is in maximum flow state.*

part from itself. So, here we examined the discreteness of the broken part. Here some queries will take place in your curious mind like why and how or in which situations we see the discrete properties in existing Universe? By not solving your queries here I tend to go further for the geometrical representation of Compton scattering. Let's have a look on the inner parts of central system and collision of massive broken part with the perfect bodies-

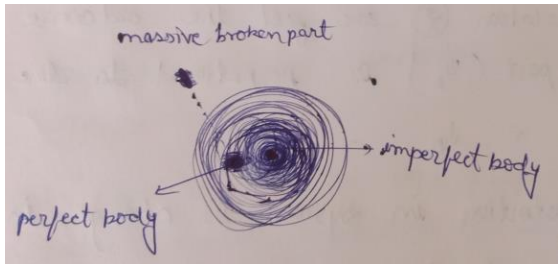

Fig.16. Broken Part Collision with Perfect Body

Here in the perfect body having low coupling scalar field, so the body need to collide with a massive broken part. This situation comes when the coupling scalar field of perfect body is the order of the massive broken part. So, the broken part will change its path after collision with the perfect body and some $\phi$-$\psi$ transformation will occur in the situation of collision. So, the broken part will lose some quantity and the perfect body's quantity will be increased after the collision.

So, from here some general conclusion comes out like in different situations (on the basis of size of broken part and comparing it with the body from which the broken part will interact) somewhere the broken part shows the discreteness and somewhere it shows uniform or continuous nature. But these properties are basically depends upon the interaction type of the broken part and body (weather it is perfect or imperfect).

From the above discussions on the behavior of broken part in different situations one beautiful query comes out in our mind that if the speed of propagation of broken part depends upon coupling with the universal scalar field then why the massive one can't travel faster than the lighter one? This depends on the properties of universal scalar field which can't be discussed in this proper article.

### 3. Some Mathematical Outcomes from the Above Geometrical Interpretations about Broken Part

From the relation (1) we get the outcome that the quantity of broken part ($\psi_b$) is proportional to the system distortion ($\delta_D$)-

$$\Psi_b \propto \delta_D$$

Then the distortion in system is change in the quantity of motion (F) according to the universal scalar field.

$$\delta_D = \frac{dF}{d\phi} \tag{11}$$

{Here F is the quantity of motion and φ is the scalar field function}

The φ is a function of various parameters but here we only include the f and $\Phi_u$ respectively the flow of scalar field and the density of scalar field where the event was occurred.

$$\delta_D = \frac{dF}{d\phi(f, \Phi u)} \tag{12}$$

So, the distortion is connected by the activities in the flow of universal scalar field.

Quantity of body also depends on the creation of the particular body and creation epoch density ($\Phi_{CB}$) and the coupling state of body (or perfectness of body)-

$$\delta_D = \frac{dF(c, \Phi_{CB}, \eta)}{d\phi(f, \Phi u)} \tag{13}$$

($\eta$=perfectness constant) & (c=creation state)

Here the perfectness constant ($\eta$) of a body depends on the coupling of the body with the universal scalar field and different types of scalar fields exist in the universe or $\eta$ is a function of $\Phi u$.

$$\eta(\Phi u)$$

So, the expression (13) becomes more complex-

$$\delta_D = \frac{dF(c, \Phi_{CB}, \eta(\Phi u))}{d\phi(f, \Phi u)} \tag{14}$$

Or the creation state includes information about the spin of the body.

Now let's have a look on the propagation speed of the broken particle from the relation (4)-

$$(v_p)_b \propto \epsilon$$

Here the coupling constant is the character of scalar fields of interaction or coupling constant depends on the density function of universal scalar field.

$$(v_p)_b \propto \epsilon(\Phi u) \tag{15}$$

The expression (15) seems not yet much clear about the propagation speed of broken particle. There should be a clear expression between the coupling and the density of scalar field.

Now by the expression (5) about the propagation speed of broken particle-

$$(v_p)_b = \frac{(\Delta \Phi u)_{covered}}{\Delta \tau}$$

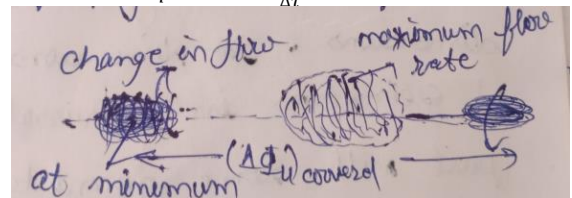

Fig.17. Flows in Broken Part

So, from here the standard definition of time comes as we measure time in sense of the flow of universal scalar field. So, there must need a universal frame of reference if we want to get the equation of motion of the various systems. Universal frame of reference has the origin on the universal singularity or from where the scalar field started to flow. So, we will get all the equations in the universal frame of reference which has the properties and connections like for each and every event the flow of time or the universal scalar field changes in it by some function.

Not going in the further explanation here I tend to give another example that can be understood by the universal scalar field flow in universal frame of reference. Edwin Hubble measured some outer broken parts as not exact as they should be or the broken parts were lost some of their quantity.

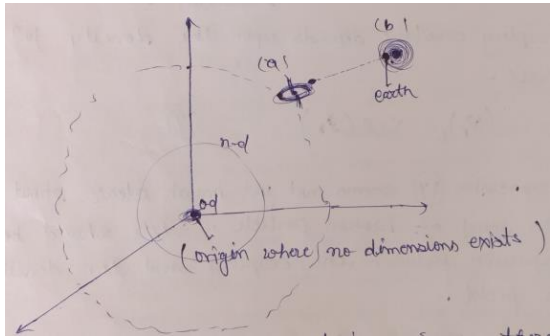

Fig.18. Doppler Effect in Universal Frame of Reference

But as the geometries formed in universe there must some dimensions. So, dimensions also change according to the formation of geometries in universal frame of reference. So, some broken parts with some quantity aparted from situation (a) and then reached to situation (b). In situation (b) the quantity of broken part is less observed than (a).

By thinking geometrically about the broken part and its path there are some central systems which changed its path by coupling state and there is some change in the universal scalar field density function. So, the behavior of the broken part is not like supposed to be.

By taking you apart from universal frame of reference here I tend to prove some another type of geometrical interpretation about the broken part formerly known as Cherenkov Effect. Here just think and draw geometrically to the situation-

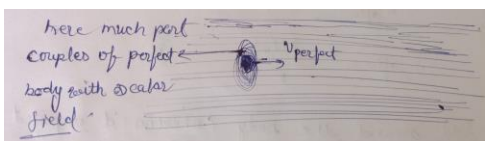

Fig.19. Coupling in Cherenkov radiation Effect

Here the phase velocity of electron or perfect body is greater than its broken part due to coupling state and the scalar field density composition. Here the body couples more than the broken part of the same with the scalar field. So, our former observations on the bodies here come out with a new proof. So, some basic definitions here come out like what is the Quantity of body exists in universe and what should be the standard measurements of the particular one? So, by not going to prove the types of the central systems and some basic definitions of some required physical quantities in this paper, I have just concluded that the speed of light or broken part can change according to various rigorous situations in universe. Here my intentions were about to prove the geometrical configuration of light and EM waves are not completed yet. So, by going straight forward to mathematical adequacies about the light or the broken particle-

From the equation (8) $\Delta\Psi_b \propto f$ the $\phi$-$\psi$ transformation comes forward in the particular equation. Here how much part converted depends on the friction and by removing proportionality with constant (k)-
$$k\Delta\Psi_b = f \qquad (16)$$
Or the $f$ is a function of the coupling constant of the particular medium which is responsible for the friction-
$$k\Delta\Psi_b = f(\epsilon_M)$$
If we take $\Psi_b$ as the full quantity of broken part then change in the quantity after the friction will be-
$$\Psi_b = (\Psi_b)_{after} + \Delta\Psi_b$$
Or $\qquad \Psi_b = (\Psi_b)_{after} + \frac{f(\epsilon_M)}{k} \qquad (17)$

Here the scalar field around the body (weather it is perfect or imperfect) will be increased by some quantity-
$$(\Phi_s)_{before} = (\Phi_s)_{after} - f(\epsilon_M)\frac{\alpha}{k} \qquad (18)$$
Or by some manipulation in (18)-
$$(\Phi_s)_{before} + f(\epsilon_M)\frac{\alpha}{k} = (\Phi_s)_{after} \qquad (19)$$

So, there is an increase in scalar field of particle when some broken part will interact with it, which is known as the basic energy-mass transformation in language of Albert. But actually the situation is more complex and beautiful then it. From here we can have the general expressions about the $\phi$-$\psi$ transformation which is continuously occurring in various central systems and the flow of universal scalar field.

From equations (17) and (19) we can conclude something-
$$\boxed{\Delta\Psi_b = -\alpha\Delta\Phi_s} \qquad (20)$$

This in general is more rigorous expression for the φ-ψ transformation.

One query absolutely taking place in your curious mind that α the conversion constant do vary in different situations in the universal frame of reference or not? Yes, for different kind of conversion α is different depending on the types and density of universal scalar field or on the system scalar field.
$$\alpha(\Phi_u, \Phi_s) \qquad (21)$$
So, α is a function of $\Phi_u$ and $\Phi_s$.

One another query also from the above generalization that α also depend upon the nature and coupling state of the body (weather it is perfect or imperfect).
$$\alpha(\Psi_s, \Phi_u, \Phi_s) \qquad (22)$$
From the above equation we can call α as the conversion constant.

So, from equation (20) we can think that what should be the basic definitions of the quantity (mass) of a particular body and energy in different sense.

## 4. Conclusion

- The concluded facts from this article totally based on the geometrical significances are-
- Speed of light is not a universal constant.
- Energy and mass relation is not the complete relation about the φ-ψ transformation.
- Each and every body has its own scalar field with its own properties.

- The universal constants like (c, G, h etc.) are different in different situations of the universe.
- The laws of physics should be defined in the universal frame of reference.
- Basic and unique definitions of the physical quantities of measurement should be needed.
- If the central systems behave in the above manner then the generalization and formation of central systems should be needed.
- Formation of various kind of geometries and characteristics of these geometries should found out.
- Universal equations of the dynamics of various types of bodies should be needed.